\documentclass[9pt]{article}
\include{macros}

\usepackage{graphicx}
\usepackage{epstopdf}
\usepackage{spconf,amsmath,amssymb,epsfig,psfrag,cite,subfigure}

\usepackage{color}
\usepackage{url}
\usepackage[margin=.94in]{geometry}

\title{Ranging Without Time Stamps Exchanging}
\name{Mohammad~Reza~Gholami, Satyam Dwivedi, Magnus Jansson, and Peter H\"andel}
\address{ACCESS Linnaeus Center, Electrical Engineering,\\
 KTH--Royal Institute of Technology, Stockholm, Sweden}

\begin{document}
\maketitle

\vspace{-5mm}
\begin{abstract}
We investigate the range estimate between two wireless nodes without time stamps exchanging.
Considering practical aspects of oscillator clocks, we
propose a new model for ranging in which the measurement errors include the
sum of two distributions, namely, uniform and  Gaussian.
We then derive an approximate maximum likelihood estimator (AMLE), which
poses a difficult global optimization problem.
To avoid the difficulty in solving the complex AMLE,
we propose a simple estimator based on the method of moments.
Numerical results show a promising performance for the proposed technique.


\end{abstract}

\vspace{-2mm}
\section{Introduction}
\vspace{-1mm}
Accurate distance estimates between two wireless nodes is a vital requirement for many applications, e.g.,
localizing the position of an unknown node \cite{Sahinoglu_2011,Sinan_book}.
Among different approaches, the range estimate based on two-way time-of-arrival (TW-TOA)
has attracted considerable attention in the literature and been proposed for practical applications, e.g.,
in IEEE 802.15.4a \cite{Sahinoglu_WAMICON}.
Traditionally, ranging based on TW-TOA involves exchanging the time stamps measured with respect to local clocks of different nodes.
Namely, a master node initiates transmitting a signal at a certain time to a slave node and receives
a response from the slave node after a delay
corresponding to the distance between two nodes and a processing time, the so-called turn-around time~\cite{Gholami_Coop_2012},
at the slave node.
The processing time can be estimated by the slave node, e.g., using a loop-back test~\cite{Sahinoglu_2011},
and then sent back to the master node.
Another approach to accurately estimate the range between two nodes is based on chirp-spread-spectrum that shows good performance in some situations~\cite{Bialer_2014,Chirp_localization_2010}.

It has been generally argued that ranging based on TW-TOA is severely affected by an imperfect clock~\cite{wang_2013,Bialer_2014,Mohammad_thesis}.
An affine model, consisting of clock offset and clock skew parameters,
is commonly considered to describe the behavior of a clock of an oscillator~\cite{chaudhari_2011_clocksynch,Sari_clock_2008,RobustClock_Localization_2011}. 
For such a model,
it is clearly
seen that the distance estimate is mainly affected by
an imperfect clock skew~\cite{gholami2013range}.
A number of studies tackled the positioning problem
in the presence of unknown clock parameters in the past few years \cite{Joint_syn_local_2010,RobustClock_Localization_2011,gholami2013range,TDOA_gholami_2013,vaghefi2013asynchronous}.

In the literature, it is commonly assumed that the clock of an
oscillator can be read continuously \cite{Erchin_2009_book,Joint_syn_local_2010}. Moreover, it is assumed that nodes are able to
communicate with each other during the ranging period.
In this study, we depart from these assumptions and consider sensor nodes as pure ranging devices.
As it is common in practice, a nominal value for the processing time is embedded in every node and once the slave node detects a signal,
it responds according to the predefined turn-around time.
Since in practice the clock of an oscillator can be read at certain times, e.g., at rising edges, we model the delay in detecting the arrival time by a uniform distribution.
We, then, model the perturbation in ranging using the sum of two distributions, i.e., uniform and Gaussian, respectively, for modeling the delay in detecting
the signal and
the time-of-arrival estimation error.
For such a model, we derive an approximate maximum likelihood estimator (AMLE), which poses a difficult global optimization problem due to nonconvexity of the AMLE objective function.
We then propose a low complexity approach based on the \emph{method of moments} (MOM) to estimate unknown parameters.
The numerical results show a promising performance of the proposed approach, specially for high signal-to-noise ratios.
In summary, the main contributions of this study are
(a) a new ranging model for practical applications;
(b) an approximate MLE (AMLE) for the proposed model;
(c) a new low complexity technique based on the MOM for estimating the unknown parameters.

%

\vspace{-4.5mm}
\section{System Model}
\label{sec:signal_model}
\vspace{-1.5mm}
Consider two master and slave nodes performing ranging using TW-TOA measurements.
For the master node, we use a time-to-digital convertor (TDC) allowing us to measure time stamp, which is not affected by imperfect clock of the master node.
Note that the TDC is a passive device; hence, it can not be used at the slave node to force the node to transmit the signal after
certain
processing delay. For details, see, e.g., \cite{Satyam_2013}.
In the literature, different models are used to describe a local clock.
Among them, the popular one
is an affine model expressed as \cite{chaudhari_2011_clocksynch,Kumar_timing_senNet,Erchin_2009_book}
\begin{align}
\label{eq:clockmodel_con}
C_s(t)=\theta_{0}+w\,t~,
\end{align}
with $C_s(t)$ as the local clock of the slave node with respect to
the perfect time.
In this model, $\theta_{0}$ and $w$ denote, respectively, the relative clock offset and the
clock skew between the slave node and the reference time $t$.
The model in \eqref{eq:clockmodel_con} assumes that the clock of an oscillator can be read continuously,
while in practice reading time happens at discrete instances, namely, at rising or falling edges.
%
%
%
%

The relation between the clock skew and the oscillator frequency offset can be seen as follows.
Let us consider the frequency $f_s$ of a slave node that deviates from nominal frequency $f_o=1/T_o$.
The common imperfections are frequency offset and phase noise phenomena. Thus, we can model the frequency $f_s$
as
\begin{align}
f_s=f_o\pm \Delta f+ \xi(t),
\end{align}
where $\Delta f$ denotes an offset and $\xi(t)$ shows the perturbation, which is assumed to be zero-mean.
Therefore, 
\begin{align}
T_s=\frac{1}{f_o\pm \Delta f+ \xi(t)}\approx T_o(1\mp \rho)+\zeta(t)
\end{align}
where $\rho\triangleq \Delta f/f_o$ is the deviation from ideal clock ($w=1\mp\rho$ is called clock skew)
and $\zeta(t)=-\xi(t)/f_o$ is known as clock jitter.
Therefore, it is observed that the period $T_s$ is a stochastic process with mean $ T_o(1\mp \rho)$.
In the rest of the paper, we assume that the jitter is small and can be neglected in the ranging process.


%

We now investigate the ranging in the presence of
clock imperfections at the slave node.
In the $k$-th round of performing TW-TOA, the master node sends a signal to the slave node
and receives a reply after a delay corresponding to the distance between the nodes and a processing delay $T^{D}_s$,
giving rise to the following model:
\begin{align}
\label{eq:TW_toa}
z_k=\frac{d}{c}+\frac{T^D_s}{2}+n_k,\quad k=1,\ldots,N
\end{align}
where $n_{k}$ is modeled by a zero-mean Gaussian random variable, i.e., $n_k\sim \mathcal{N}(0,\sigma^2)$~\cite{Sahinoglu_2011,Gholami_ICC2011}, and
$c$ is the speed of propagation,~$d$
is the Euclidian distance between two nodes.
We now model the delay $T^D_s$ as follows.
In practice, a nominal value for the delay is set into every node, say, $T^D=DT_o$, where $D$ is an integer.
It means that the slave node replies the detected signal after $wDT_o$.
Therefore, the total delay at slave node can be modeled by
\begin{align}
\label{eq:turn_around}
T^D_s=wDT_o+\epsilon_k
\end{align}
where $\epsilon_k$ determines the delay in detecting the signal presence.
In fact, the arrived signal at the slave node may be detected after
$\epsilon_k$, i.e., at the first rising edge of the clock.
A natural way to model $\epsilon_k$ is to employ a uniform distribution
$\epsilon_k\sim \mathcal{U}(0,wT_o)$. In fact, for high-signal-to-noise ratios (SNRs), with high probability time-of-arrival detection
happens in the period that signal arrives.


Replacing \eqref{eq:turn_around} into \eqref{eq:TW_toa}, we obtain the following model for ranging via TW-TOA:
\begin{align}
\label{eq:TW_toa_finla}
z_k=\frac{d}{c}+\frac{wDT_o}{2}+\frac{\epsilon_k}{2}+n_k\quad k=1,\ldots,N.
\end{align}
Clearly, it is seen that the perturbation $\epsilon_k/2+n_k$ has nonzero mean $wT_o/4$, which depends on unknown clock skew.
For large $D$, we may successfully neglect $\epsilon_k$ in \eqref{eq:TW_toa_finla}, especially for tiny $T_o$,
and arrive at the traditional model of ranging considered in the literature. But, in general,
the delay in detecting the arrival signal needs to be taken into account to have a more accurate model.
In addition, a small value for the processing delay in slave node, small $D$, is preferable for some applications, e.g., for fast ranging.
%
\vspace{-5mm}
\subsection{Maximum Likelihood Estimator}
\begin{figure*}
\vspace{-5mm}
\centering
\subfigure[]{
 \psfrag{d}[cc][][.8]{$d$}
 \psfrag{f}[cc][][.8]{$w$}
 \psfrag{ML cost function}[cc][][.7]{AMLE cost function}
  \includegraphics[width=53mm]{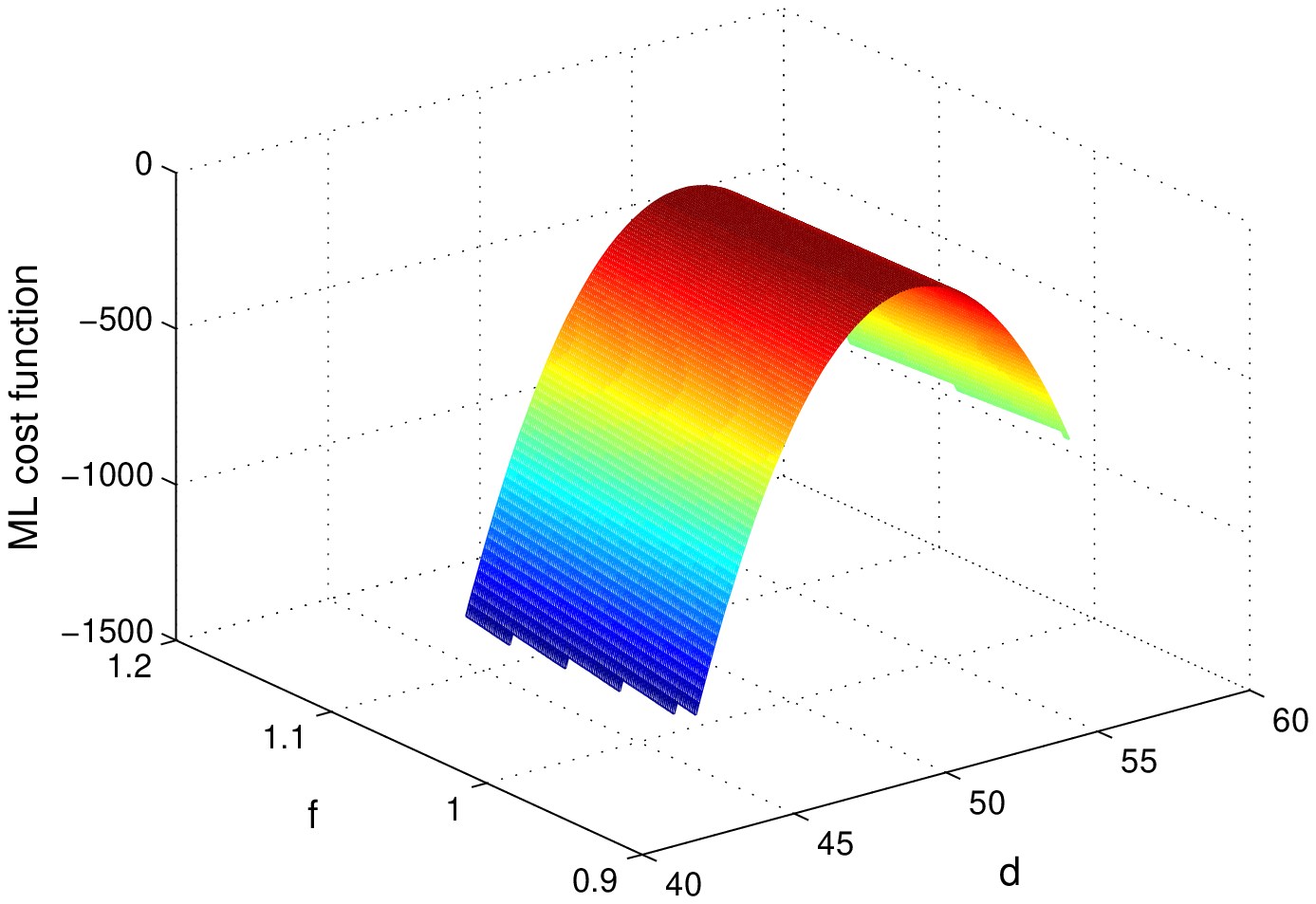}
 \label{eq:fixed_sig}
  }
  \subfigure[]{
  \psfrag{d}[cc][][.8]{$d$}
 \psfrag{sigma}[cc][][.8]{$\sigma$}
 \psfrag{ML cost function}[cc][][.7]{AMLE cost function}
   \includegraphics[width=53mm]{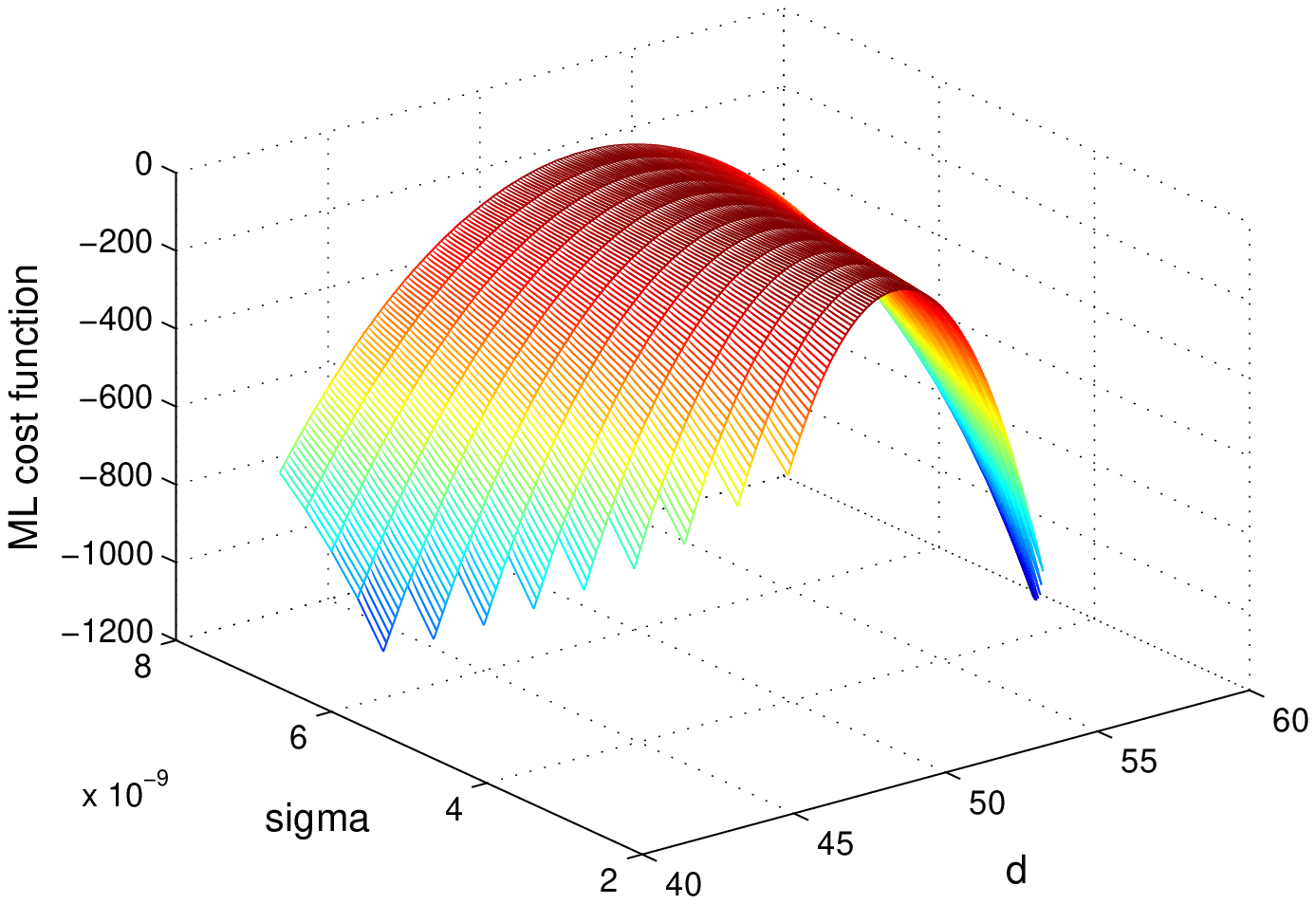}
   \label{eq:fixed_f}
  }
  \subfigure[]{
  \psfrag{f}[cc][][.8]{$w$}
 \psfrag{sigma}[cc][][.8]{$\sigma$}
 \psfrag{ML cost function}[cc][][.7]{AMLE cost function}
  \includegraphics[width=53mm]{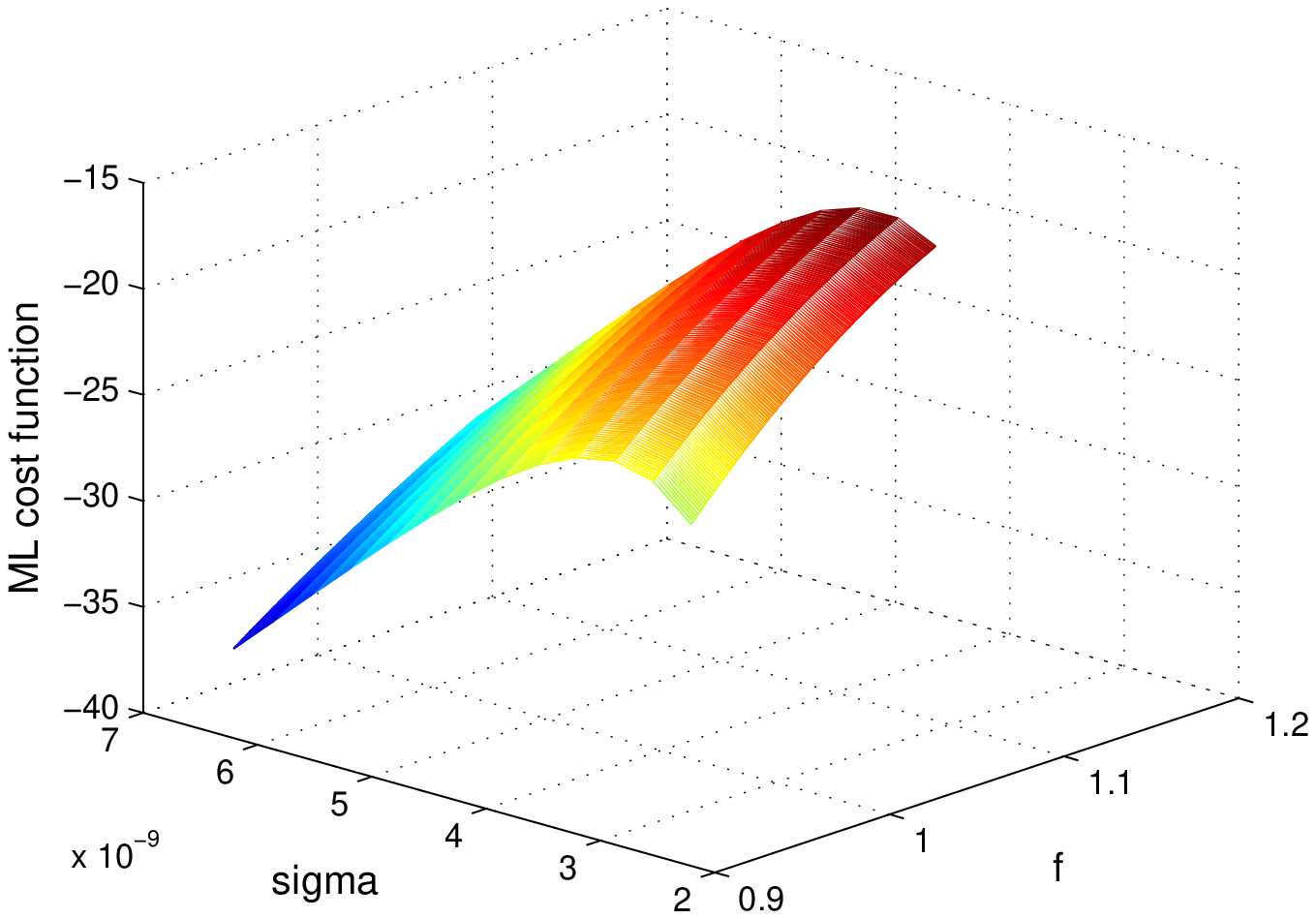}
  \label{eq:fixed_d}
  }
  \vspace{-3mm}
   \caption{AMLE cost function for \subref{eq:fixed_sig} $d$ and $w$ for fixed $\sigma=1/c$, \subref{eq:fixed_f}  $d$ and $\sigma$ for fixed $w=1.0001$, and
 \subref{eq:fixed_d} $w$ and $\sigma$ for fixed $d=50$.}
 \label{eq:cost_plot}
 \vspace{-5mm}
  \end{figure*}

We consider the vector of measurement $\boldsymbol{z}\triangleq [z_1,\ldots,z_N]^T$
and assume that $\epsilon_k$ and $n_k$ are independent. In addition, it is assumed
that $n_k ( \text{or}~ \epsilon_k)$ and $n_{\ell} ( \text{or}~ \epsilon_{\ell})$ are independent for $k\neq \ell$.
Thus, the probability density function (pdf) of the measurement vector $\boldsymbol{z}$
can be calculated as
\begin{align}
\label{eq:ML}
&p_{\boldsymbol{Z}}(\boldsymbol{z};\boldsymbol{\theta})
=\prod_{k=1}^N  \int_{x=0}^{wT_0} p_{{Z}_k}({z}_k|x,~\boldsymbol{\theta})p_{\epsilon_k}(x)dx\nonumber\\
&=\prod_{k=1}^N \int_{x=0}^{wT_0} \frac{1}{\sqrt{2\pi}\sigma wT_0}\exp{\left(-\frac{(\alpha_k-x/2)^2}{2\sigma^2}\right)}dx
\end{align}
where $\alpha_k\triangleq z_k-d/c-wDT_0/2$
and $\boldsymbol{\theta}\triangleq [d,w,\sigma]$.
The integral in \eqref{eq:ML} has no closed-form expression.
For any $x\in \mathbb{R}$, we instead use the following approximation~\cite{Gauss_approx_2012}:
\begin{align*}
\int_{0}^x \exp(-\pi t^2)dt\approx \frac{1}{2}\tanh\left(\frac{39x}{2}-\frac{111}{2}\arctan\left(\frac{35x}{111}\right)\right).
\end{align*}
Using the above approximation, we obtain
\begin{align*}
 &\int_{x=0}^{wT_0} \frac{1}{\sqrt{2\pi\sigma^2}wT_0}\exp{\left(-\frac{(\alpha_k-x/2)^2}{2\sigma^2}\right)}dx\nonumber\\
 &=\frac{2}{wT_0}\Bigg[\int_{0}^{\frac{\alpha_k}{\sqrt{2\pi}\sigma}}\exp{(-\pi t^2)}dt
 -\int_{0}^{\frac{\alpha_k-wT_0/2}{\sqrt{2\pi}\sigma}}\exp{(-\pi t^2)}dt\Bigg]\nonumber\\
 &=\frac{1}{wT_0}\Bigg[\tanh\left(\frac{39\alpha_k}{2\sqrt{2\pi\sigma}}-\frac{111}{2}\arctan\left(\frac{35\alpha_k}{111\sqrt{2\pi\sigma}}\right)\right)\nonumber\\
 &\qquad-\tanh\left(\frac{39\beta_k}{2\sqrt{2\pi\sigma}}-\frac{111}{2}\arctan\left(\frac{35\beta_k}{111\sqrt{2\pi\sigma}}\right)\right)\Bigg],
\end{align*}
where $\beta_k=\alpha_k-wT_0/2$. Hence, we obtain the following approximation for the pdf of measurements in \eqref{eq:ML}:
\begin{align*}
&p_{\boldsymbol{Z}}(\boldsymbol{z};\boldsymbol{\theta})=(wT_0)^{-N}\prod_{k=1}^N\nonumber\\
&\Bigg[\tanh\left(\frac{39\alpha_k}{2\sqrt{2\pi\sigma}}-\frac{111}{2}\arctan\left(\frac{35\alpha_k}{111\sqrt{2\pi\sigma}}\right)\right)\nonumber\\
 &~~-\tanh\left(\frac{39\beta_k}{2\sqrt{2\pi\sigma}}-\frac{111}{2}\arctan\left(\frac{35\beta_k}{111\sqrt{2\pi\sigma}}\right)\right)\Bigg].
\end{align*}
We now obtain the AMLE expression by solving an optimization problem, expression \eqref{eq:amle_expression} shown at the top of the page.
\begin{figure*}[!t]
\begin{align}
\label{eq:amle_expression}
\mathop{\mathrm{maximize}}\limits_{\sigma;~w;~d} -N(wT_0)+\sum_{k=1}^N \log \Bigg[&\tanh\left(\frac{39\alpha_k}{2\sqrt{2\pi\sigma}}-\frac{111}{2}\arctan(\frac{35\alpha_k}{111\sqrt{2\pi\sigma}}\right)
 -\tanh\left(\frac{39\beta_k}{2\sqrt{2\pi\sigma}}-\frac{111}{2}\arctan(\frac{35\beta_k}{111\sqrt{2\pi\sigma}}\right)\Bigg]
\end{align}
\end{figure*}
It is observed that the optimal estimator poses a difficult global optimization problem.
To get some feeling about the shape of the objective function, we fix one parameter and plot
the value of the objective function over two other parameters in Fig. \ref{eq:cost_plot}.
As it is observed, the objective function for a fixed parameter has many maxima and it may also be discontinuous for some values.
In the coming section, we propose a simple estimator based on the method of moments (MOM).

\vspace{-3mm}
\subsection{A Low Complexity Estimator}
We first consider the relations between the unknown parameters and the following statistics:
\begin{align}
\mu_{1}&\triangleq \mathbb{E} z_k= \frac{d}{c}+\left(\frac{D}{2}+\frac{1}{4}\right)wT_{0}\label{eq:mom_1}\\
\mu_{2}& \triangleq \mathbb{E} (z_k-\mu_1)^2=\sigma^2+\frac{(wT_0)^2}{48}\label{eq:mom_2}\\
\mu_{4}& \triangleq \mathbb{E} (z_k-\mu_1)^4=3\sigma^4+\sigma^2\frac{(wT_0)^2}{8}+\frac{(wT_0)^4}{1280}\label{eq:mom_3}.
\end{align}
The statistics $\mu_1$, $\mu_2$, and $\mu_4$ can be approximated by the means of
ensemble averaging as follows:
\begin{align*}
&\mu_{1}\approx \frac{\sum_{k=1}^N z_k}{N}=S_1,\\
& \mu_{2} \approx  \frac{\sum_{k=1}^N(z_k-S_1)^2}{N}=S_2\\
&\mu_{4} \approx \frac{\sum_{k=1}^N(z_k-S_1)^4}{N}=S_4.
\end{align*}
From \eqref{eq:mom_2} and \eqref{eq:mom_3}, we obtain
\begin{align}
\label{eq:estimator_to_derived}
a(wT_o)^4=\mu_4-3\mu_2^2
\end{align}
where $a\triangleq 1/1280-3/48^2$. 
Hence, an estimate of the clock skew is obtained as
\begin{align}
\label{eq:final_estiamte}
\widehat{w}=\frac{1}{T_o}\left(\left|\frac{S_4-3S_2^2}{a}\right|\right)^{1/4}.
\end{align}
We use the absolute value in \eqref{eq:final_estiamte} to prevent ambiguity due to noise in estimating the clock skew.

Considering $w^K=(1+\rho)^K\approx 1+K\rho$, we can also estimate the clock skew from \eqref{eq:estimator_to_derived} as
\begin{align}
\label{eq:final_estiamte_2}
\widetilde{w}=1+\frac{aT_o^4-S_4+3S_2^2}{4aT^4_o}.
\end{align}
We can now estimate the distance and the variance $\sigma^2$ as
\begin{align}
&\widehat{d}=c\left(S_1-\left(\frac{D}{2}+\frac{1}{4}\right)\widetilde{w}T_{0}\right),\label{eq:dis_estimate_final}\\
& \widehat{\sigma}^2=\left|S_2-\frac{(\widetilde{w}T_0)^2}{48}\right|.
\end{align}
The mean of the estimator proposed in \eqref{eq:final_estiamte_2} is
\begin{align}
\label{eq:rho_est_mean}
 \mathbb{E} \widetilde{w}&= 1+\frac{aT_o^4-\mathbb{E}S_4+3\mathbb{E}S_2^2}{4aT^4_o}.
\end{align}
Using the law of large numbers, we have
\begin{align*}
\frac{1}{N}\sum_{k=1}^Nz_k \rightarrow \mu_1=\frac{d}{c}+\left(\frac{D}{2}+\frac{1}{4}\right)wT_{0}, \quad N\rightarrow \infty.
\end{align*}
We now compute the expectations on the right-hand side of \eqref{eq:rho_est_mean} as follows.
\begin{align*}
\mathbb{E}S_4
&=\mathbb{E} \frac{1}{N}\sum_{k=1}^N(\tilde{z}_k^4-4\tilde{z}_k^3\tilde{S}_1+6\tilde{z}_k^2\tilde{S}^2_1-4\tilde{z}_k\tilde{S}^3_1+\tilde{S}_1^4)
\end{align*}
where $\tilde{z}_k \triangleq z_k-\mu_1$ and $\tilde{S}_1=1/N\sum_{\ell}^N(z_k-\mu_1)$.
Since $\tilde{z}_k$ are iid, we have
\begin{align*}
&\mathbb{E}\tilde{z}_k^3\tilde{S}_1=\frac{1}{N}\mathbb{E} \tilde{z}_k^4,\quad\mathbb{E}\tilde{z}_k\tilde{S}^3_1=\frac{\mathbb{E} \tilde{z}_k^4}{N^3}+\frac{3(N-1)}{N^3}(\mathbb{E}\tilde{z}_k^2)^2 \nonumber\\
&\mathbb{E}\tilde{z}_k^2\tilde{S}^2_1=\frac{\mathbb{E} \tilde{z}_k^4+(N-1)\mathbb{E}\tilde{z}^2_k}{N^2},~~
\mathbb{E}\tilde{S}^4_1=\frac{\mathbb{E} \tilde{z}_k^4+(N-1)\mathbb{E}\tilde{z}^2_k}{N^3}
\end{align*}
Hence, 
\begin{align}
\mathbb{E}S_4&=\left(1+\frac{2N-3}{N^3}\right)\mathbb{E}\tilde{z}_k^4+\frac{6(N-1)^2}{N^3}(\mathbb{E}\tilde{z}_k^2)^2\nonumber\\
&=\mathbb{E}\tilde{z}_k^4+O(1/N)
\end{align}
Likewise, we can show that
\begin{align}
\mathbb{E}S_2^2&=(\mathbb{E} \tilde{z}_k^2)^2+O(1/N).
\end{align}
Considering $\mathbb{E}\tilde{z}_k^4=3\sigma^4+\sigma^2{(wT_0)^2}/{8}+{(wT_0)^4}/{1280}$ and $\mathbb{E}\tilde{z}_k^2=\sigma^2+(wT_0)^2/48$,
it is verified that
\begin{align}
\label{eq:rho_est_mean}
 \mathbb{E} \tilde{w}&= 1+\frac{aT_o^4-a(wT_0)^4}{4aT^4_o}+O(1/N)\nonumber\\
  &\approx 1+\frac{aT_o^4-a(wT_0)^4}{4aT^4_o}\approx w,\quad N\gg 1,
\end{align}
meaning that the estimator is unbiased for large number of samples.
\begin{figure}
\vspace{-5mm}
 \centering
\subfigure[]{
 \psfrag{xlabel}[cc][][.8]{Standard deviation of noise, $\sigma$}
 \psfrag{ylabel}[cc][][.8]{Mean of the distance estimate [m]}
 \psfrag{Proposed}[cc][][.45]{Proposed}
 \psfrag{Traditional}[cc][][.45]{Traditional}
 \psfrag{Counter-based}[cc][][.45]{Counter-based}
 \includegraphics[width=60mm]{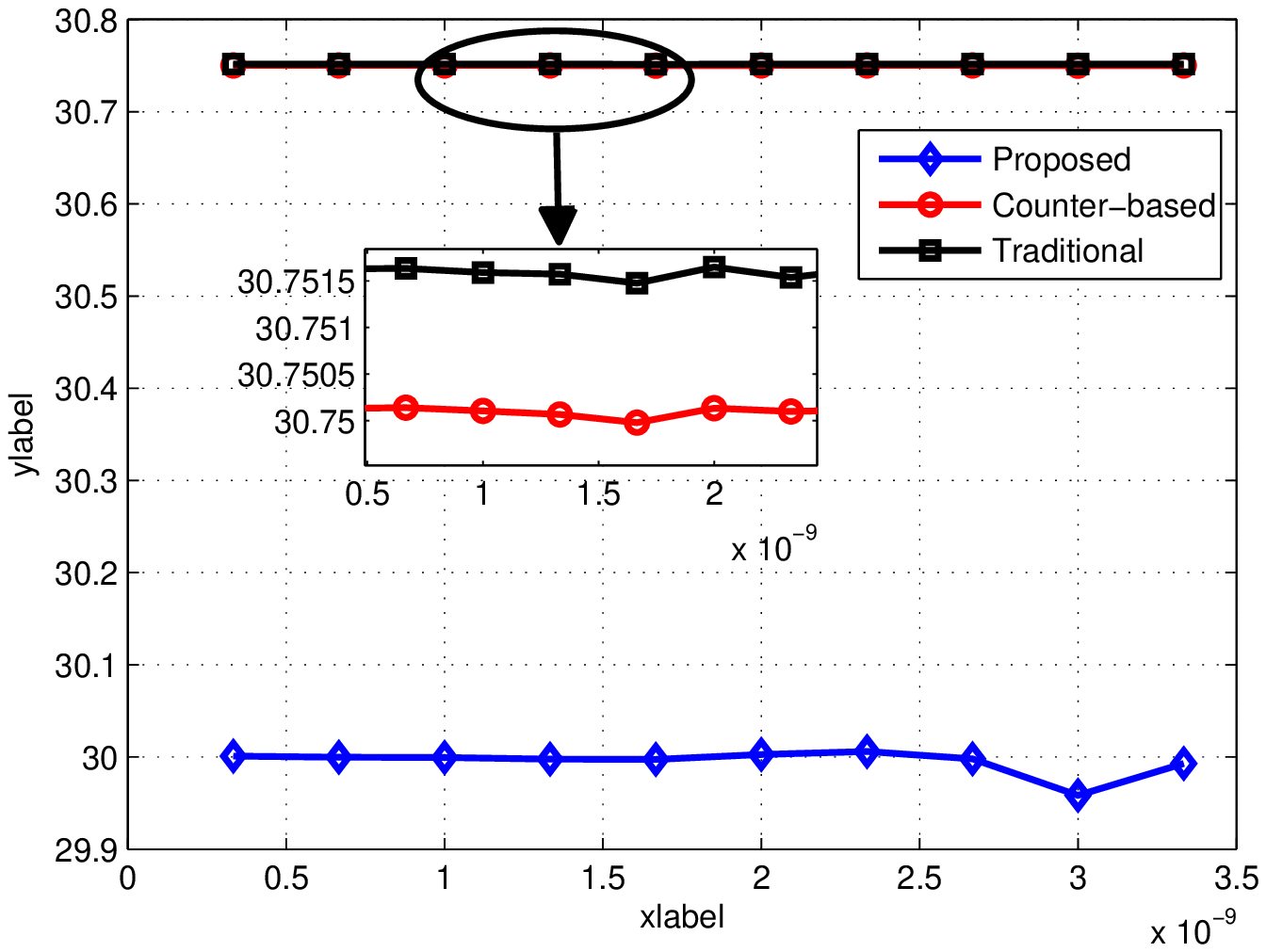}
\label{fig:dis_est}}
\subfigure[]{
 \psfrag{xlabel}[cc][][.8]{Standard deviation of noise, $\sigma$}
 \psfrag{ylabel}[cc][][.8]{RMSE [m]}
 \psfrag{Proposed}[cc][][.45]{Proposed}
 \psfrag{Traditional}[cc][][.45]{Traditional}
 \psfrag{Counter-base}[cc][][.45]{Counter-based}
  \includegraphics[width=60mm]{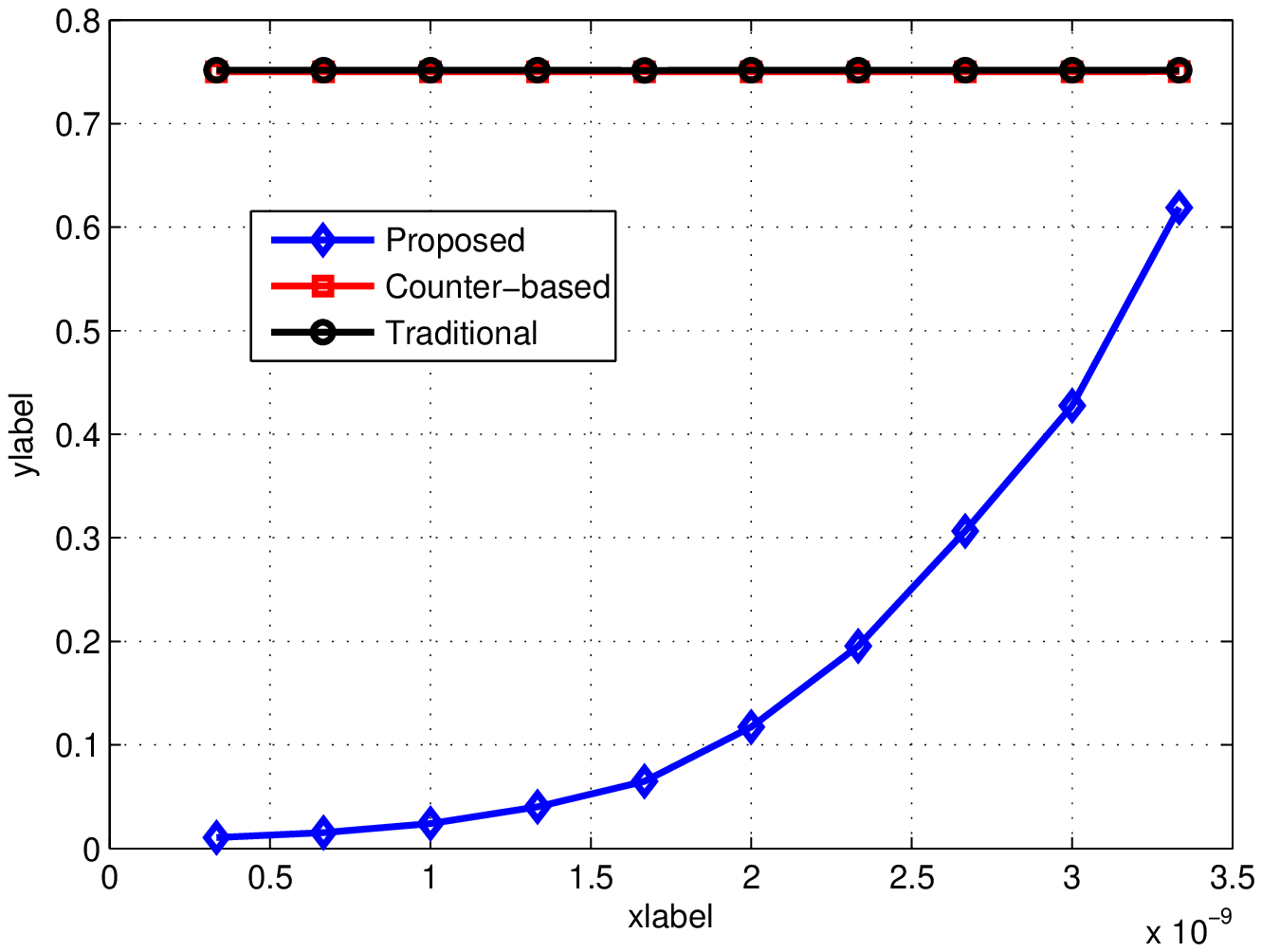}
\label{fig:rmse_es}}
\vspace{-5mm}
\caption{Comparison between different approaches,  \subref{fig:dis_est} the mean of distance estimate and \subref{fig:rmse_es} the RMSE of the estimate.}
 \label{fig:comparison}
\end{figure}
%
\section{Simulation results}
We compare the proposed technique with the traditional approach without clock skew compensation
and a technique based on approximating the turn-around time at the slave node via loop-back test and then sending back the estimate to the master node.
In the simulation, we use $d=30$ [m], $f_o=100$ MHz. We consider $\rho=0.0001$, which corresponds to a frequency offset equal to $10$ kHz.
In the simulation, we set $D=10$ and we collect TW-TOA measurements for 10 ms. 
To obtain the results, we run the algorithms for 1000 realizations of noise.

In Fig.\,\ref{fig:comparison}, we plot the mean of the distance estimate and the root-mean-square error (RMSE) of the estimate for
different approaches. As it is observed, the proposed technique shows a considerable gain, especially for
high SNRs.

\begin{figure}
\vspace{-5mm}
 \centering
 \psfrag{xlabel}[cc][][.8]{N}
 \psfrag{ylabel}[cc][][.8]{$\mathbb{E} \widehat{d}$ [m]}
 \psfrag{Theoretical}[cc][][.8]{Theoretical}
 \psfrag{Actual}[cc][][.8]{Actual}
  \includegraphics[width=70mm]{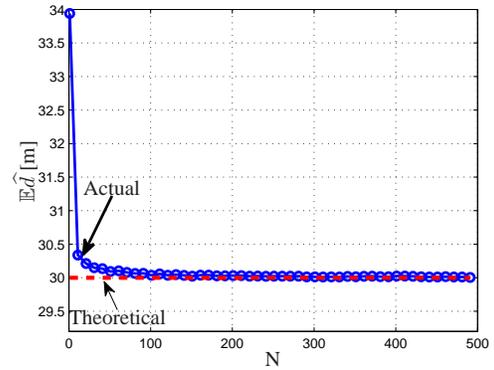}
  \vspace{-5mm}
\caption{The mean of the distance estimate versus the number of samples $N$.}
 \label{fig:num_sam}
\end{figure}

Fig.\ref{fig:num_sam} shows the mean of the estimate versus the number of
distance estimates, i.e., Eq. \eqref{eq:dis_estimate_final}, $N$ for fixed $\sigma=0.1/c$. As it is observed, after a sufficient number of samples, e.g., 100 corresponding to $1$ micro second,
the estimate is very close to the theoretical value.


\section{Conclusions}
In this paper, we have studied the distance estimate between two nodes using TW-TOA.
Considering practical aspects, we have modeled the range estimate using the sum of two distributions, uniform and
Gaussian. We showed that the optimal estimator is complex and not easy to solve.
Then, we proposed a low complexity technique based on the method of moments.
Numerical results show a promising performance for the proposed approach, especially for high signal-to-noise ratios.

\bibliographystyle{IEEEtran}
\bibliography{../../Reference/Ref}

\end{document}